# Magnetic fingerprint in a ferromagnetic wire: Spin torque diode effect and induction of the DC voltage spectrum inherent in the wire under application for RF current

A. Yamaguchi[1], T. Ono[2], Y. Suzuki[3], S. Yuasa[4], and H. Miyajima[1]

[1]Department of Physics, Faculty of Science and Technology, Keio University, 3-14-1 Hiyoshi,

Yokohama, Kanagawa 223-8522, Japan

[2]Institute for Chemical Research, Kyoto University, Gokasho, Uji, Kyoto 611-0011, Japan

[3]Graduate School of Engineering and Science, Osaka University, Machikaneyama 1-3,

Toyonaka, Osaka 560-8531, Japan

[4]Nanoelectronics Research Institute, National Institute of Advanced Industrial Science and

Technology, Tsukuba 305-8568, Japan

**[Abstract]**

We report the rectifying effect of a constant-wave radio frequency (RF) current by a magnetic domain wall (DW) on a single-layered ferromagnetic wire. A direct-current (DC) voltage is generated by the spin torque diode effect, which is a consequence of magnetoresistance oscillation due to the resonant spin wave excitation induced by the spin-polarized RF current. The DC voltage spectrum strongly depends on the internal spin structure in the DW, which corresponds to the magnetic fingerprint of the spin structure in the ferromagnetic wire.

A magnetic domain wall (DW) is the magnetization spatial transition region between magnetic domains, in which the magnetic moments undergo reorientation. Most of the magnetic changes under the action of weak and moderate magnetic fields occur at DWs and understanding of DW behavior is hence one of the essentials in describing the magnetizing process. There are many studies on planer displacement of rigid DWs under the action of the magnetic field [1 – 3] and electrical current [4 – 13]. In general, the DW is subjected to magnetic potential energy in a ferromagnet. The potential energy varies throughout the ferromagnet since some defects existing in the material provide minima of the potential energy, while the region of inhomogeneous microstress, associated with dislocations, provides either energy minima or maxima depending on the sign of the stress and magnetostriction coefficient [1 – 3]. The fingerprint appropriate for the random potential in magnetic materials has been studied from the viewpoint of the Barkhausen effect in Fe-Ni-Co alloy [14] and from the magnetic field dependence of conductivity fluctuations in a variety of mesoscopic systems [15, 16, 17]. Recently, Vila *et al*. have reported that DWs make a reproducible spin contribution to the universal conductance fluctuation and that they also lead to hysteric behavior with irreproducible fluctuations observed in the minor loop [17]. The shape of the potential is irregular in the space and acts on the internal spin structure of the DW in the absence of a magnetic field.

When a spin-polarized electrical current flows through some spin structure with variation such as ferromagnetic multilayers [18 – 21], the DW [4 – 9], or a magnetic vortex [22], magnetization switching and magnetic excitation are induced by the spin-transfer effect [23, 24], which is a consequence of the spin angular momentum transfer occurring the interaction between the spin-polarized current and the magnetic moment. One of the interesting properties is the spin-torque diode effect found in the magnetic tunnel junction (MTJ) [24], the spin-valve giant magnetoresistive device (SV-GMR) [25], and single-layered ferromagnetic wires [27]. It should be noted that direct-current (DC) voltage is produced when a radio-frequency (RF) current flows through a nanometer-scale MTJ, a SV-GMR, or a ferromagnetic submicron wire. This DC voltage generation originates in the magnetoresistance oscillation due to the ferromagnetic resonance (FMR) generated by the spin-transfer effects [25, 26]. As reported in a previous paper [27], the spin torque diode effect in the single-layered ferromagnetic wire is the consequence of the magnetoresistance oscillation generated by the spin-transfer torque. The torque is a combined effect of the spatial non-uniformity of the magnetization and the flow of the spin-polarized current.   In the other words, the spin torque diode effect is highly sensitive to the spin structure in nano-scale magnets.

In this study, we present the spin torque diode effect in a single-layered ferromagnetic wire without an external magnetic field. The DC voltage spectrum depends on the internal spin structure at the DW, which corresponds to the magnetic fingerprint reflected by the variation of the potential energy in the wire.

Two kinds of wire, boomerang-shaped and semicircular-shaped ferromagnetic wires of $Ni_{81}Fe_{19}$, were fabricated on MgO substrates using electron beam lithography and the liftoff method. Figure 1(a) shows an optical micrograph of the boomerang-shaped wire and the present RF electric circuit.   The size and shape of the wires prepared in this study are listed in Table 1. The measurement system is almost same as that described in the previous paper [27]. A microwave probe is connected by the wire, and the bias tee circuit detects the DC voltage difference induced by the RF current flowing through the wire, as schematically illustrated in Fig. 1 (a). A vector network analyzer (frequency range: 45 MHz – 67 GHz) or a signal generator (frequency range: 10 MHz – 40 GHz) injects the RF current into the wires. All experiments are performed at room temperature in the absence of an external magnetic field.

In the soft ferromagnetic $Ni_{81}Fe_{19}$ narrow wire, the magnetization is directed along the wire axis owing to the presence of strong uniaxial shape anisotropy. Therefore, we are able to introduce or remove a single DW in the following manner: A DW is introduced (removed) by applying a magnetic field, $H$=1 kOe, at an angle of $\theta = 90°$ ($\theta = 0°$) so as to saturate the magnetization and decrease the field to zero. The magnetic structures of the head-to-head DW and no DW in the wire calculated by a micromagnetics simulator (OOMMF) [28] are shown in Fig. 1 (b) and (c), respectively. The parameters used for the calculation are as follows: a unit cell size of 5 nm × 5 nm with a constant thickness of 50 nm, a magnetization of 1.08 T, and a damping constant of $\alpha = 0.1$. The size of the simulation model is same as sample SC1 apart from the wire length.

The dynamical properties of the DW magnetization are probed using the vector network analyzer in the microwave reflection mode. In order to remove the effect of cables, on-wafer calibration is performed using the open-short-load calibration procedure [29]. We calibrate reflection signal $S_{11}$ of sample SC1 with the DW, and measure signal $S_{11}$ after the DW is removed in the previously described manner. In Fig. 1 (d), the red dashed line and the black solid line correspond to $S_{11}$ with and without a DW, respectively. The eigenmodes of the magnetization in the wire with and without a DW are determined by free damping applying a pulsed magnetic field along the z-direction (perpendicular to the plane). After this excitation, the magnetization in the wire performs damped free oscillations. Figures 1(e) and 1(f) show the time variation of the z-componet of the magnetization and the Fourier spectrum, respectively. As shown in Fig. 1(f), the wire without a DW has an intrinsic frequency of 9 GHz, while the wire with a DW takes a frequency of 1, 2, and 8 GHz. The frequency of the magnetization oscillation in the wire with a DW is smaller than that in the wire without a DW. The dips in $S_{11}$ of the wire without a DW at frequency of 9.5 and 10.4 GHz correspond to the magnetization oscillation in the wire with and without a DW, respectively. The decrease in $S_{11}$ of the wire without a DW in the range lower than 5 GHz also corresponds to the magnetization oscillation calculated in the wire with a DW.

Figure 2 shows the RF current frequency (RF power: -5 dBm) dependence of the DC voltage difference due to the spin torque diode effect in the absence of a magnetic field. The black dashed line and the red solid line correspond to the spectrum without a DW and with a DW, respectively. As shown in Fig. 2, the observed resonance spectrum of the wire with a DW is more complicated than that of the wire without a DW. This indicates that the complicated resonance spectrum is a consequence of the spatial non-uniformity of the magnetization due to the existence of a DW. In other words, the effective magnetic field is homogenous in the wire without a DW, but the effective field depends on the position of the wire with a DW due to the internal spin structure. The spatial inhomogeneity of the effective field produces the complicated spectrum. That is, the spin torque diode signal shows a pattern inherent in each wire, or a magnetic fingerprint, reflecting the spatial variation of the spin structure with high sensitivity compared with $S_{11}$ measurement.

Figure 3 shows the RF power and frequency dependences of the induced DC voltage difference for samples SC1 and SC2 with a DW. The amplitude of the DC voltage spectrum increases with increasing RF current, but the spectrum structure is conserved. This indicates that the spin-transfer torque or the Joule heating due to the RF current is not as large as the magnetic spatial structure disarranged. That is, there is spectral reproducibility [14 – 17].

We repeated erasing the DW, introducing the DW again, and measuring the rectifying effect of the RF current in the wire in the absence of an external magnetic field. The DC voltage spectra of samples SC2 and SC3 are shown in Fig. 4 (a) and (b), respectively, in which the number following the symbol # means the number of measurement times. As is seen in Fig. 4 (a), three kinds of spectrum are observed; the spectra for #2, #3, #4, and #6 are the same as each other but different from those of #1 and #5, while the spectra for #1 and #6 are the same but different from those of #2, #3, #4, and #5, as shown in Fig. 4 (b). After the DW is erased and introduced, a small change in the spin configuration in the wire slightly modifies the magnetic fingerprint of the DW. This is attributed to the internal spin structure within the DW influenced by the pinning potential associated with certain defects such as dislocations, impurities, the micro-structural inhomogeneties, or edge roughness in the wire. As shown in Fig. 4, after the DW is reintroduced into the wire, a new spin configuration including the DW, which is influenced by the random pinning potential sites in the wire, results in a change in the magnetic fingerprint. As shown in Fig. 3, the spin structure once made is unchanged by the spin-transfer torque or the thermal fluctuation caused by Joule heating due to the RF current that is not as large as the magnetic spatial structure disarranged. As a result, the magnetic fingerprint due to the spin torque diode effect provides information of the variation of the pinning potential sites in the wire.

In conclusion, the DC voltage is induced by the spin torque diode effect in a single-layered ferromagnetic wire without an external magnetic field. The frequency spectrum of the DC voltage is influenced by the microscopic spin configuration inherent in the random pinning sites in the wire, which corresponds to the magnetic fingerprint. The magnetic fingerprint is a highly sensitive measurement of spin structure in nano-scale magnets.

The present work was partly supported by MEXT Grants-in-Aid for Scientific Research in Priority Areas, JSPS Grants-in-Aid for Scientific Research, an Industrial Technology Research Grant Program in '05 from NEDO of Japan, and the Keio Leading-edge Laboratory of Science and Technology project 2006.

**[Figure captions]**

Figure1

(a) The radio-frequency (RF) electric circuit and the optical micrograph (top view) of the system consisting of electrodes and a magnetic nanowire. The magnetic field is applied in the substrate plane at an angle of $\theta$ to the longitudinal axis of the nanowire. The direct-current (DC) voltage across the wire is measured by using a bias tee. The in-plane magnetization distribution (b) with and (c) without a domain wall (DW) at the bottom of the wire loop visualized by OOMMF. (d) Reflection signal $S_{11}$ as a function of the RF current frequency. (e) The temporal evolution of $M_z$ during free damping. (f) The averaged power spectrum for $M_z$.

Figure 2

The DC voltage induced by the spin torque diode effect as a function of the RF current frequency. The red solid line and the black dashed line correspond to the DC voltage spectrum with and without DW, respectively.

Figure 3

The DC voltage spectra for SC1 and SC2 as a function of RF power during the existence of the DW.

Figure 4

The magnetic fingerprint of (a) SC2 and (b) SC3. Each spectrum is measured after the DW is erased and then introduced again.

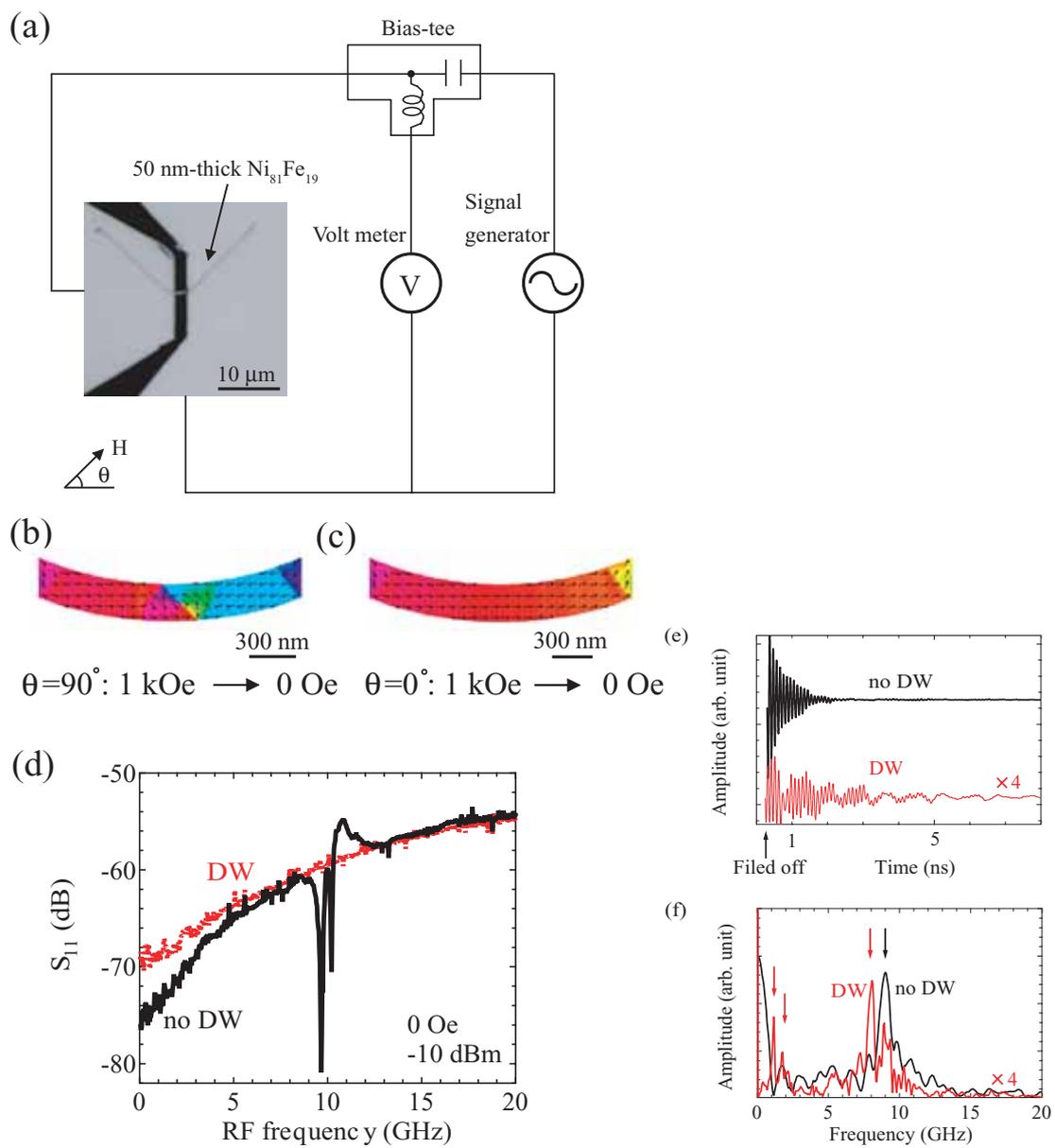

Fig. 1

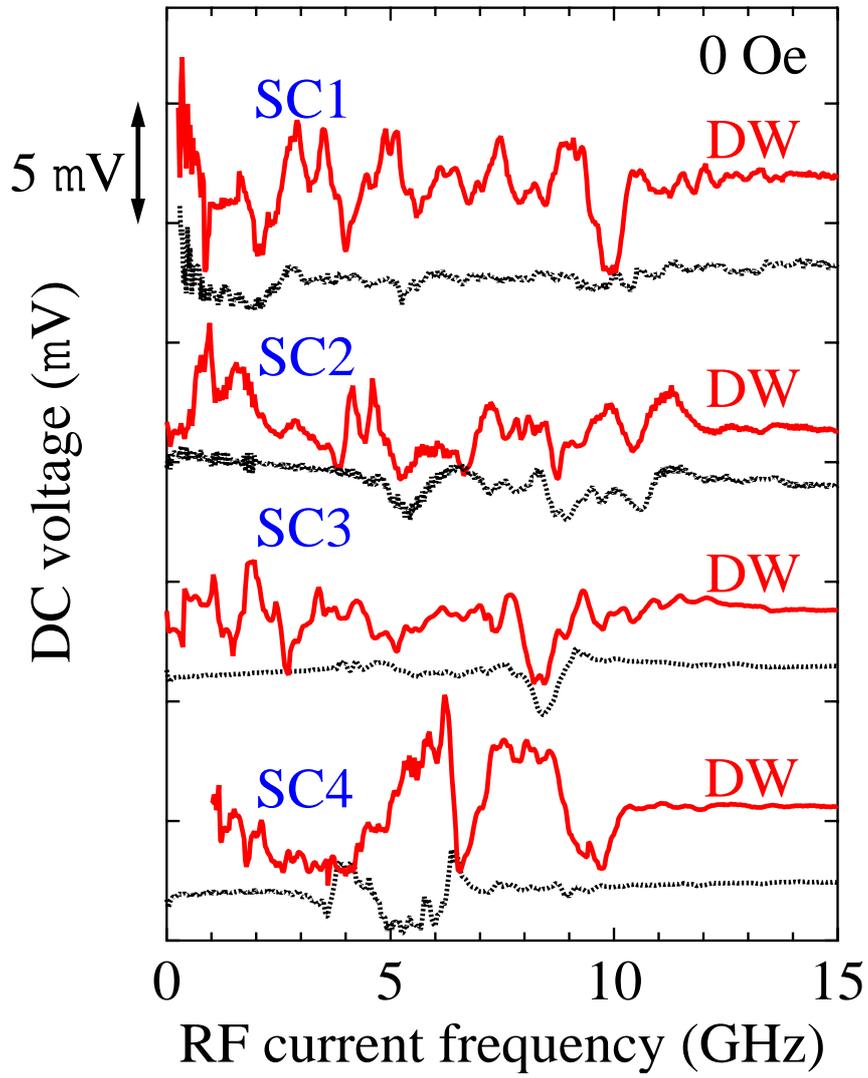

Fig. 2

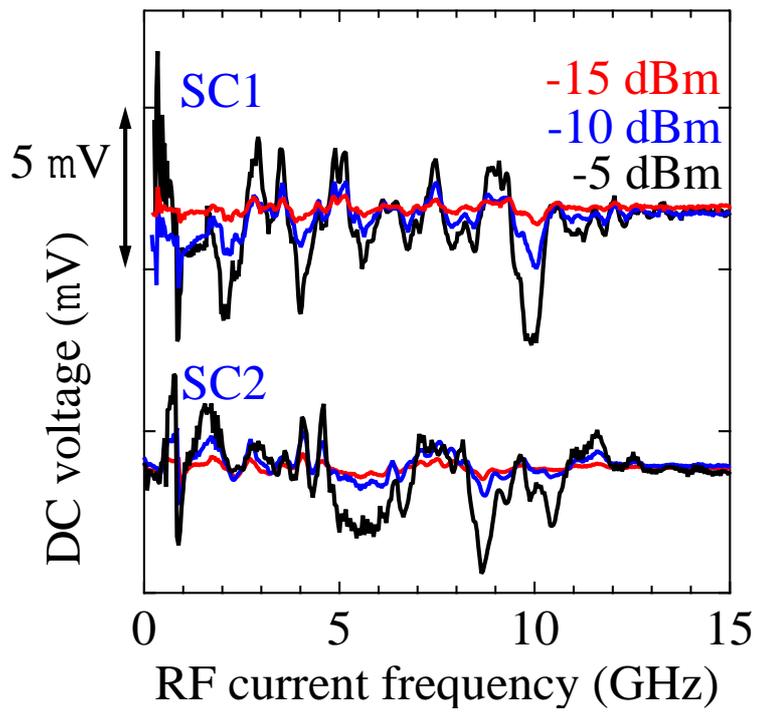

Fig. 3

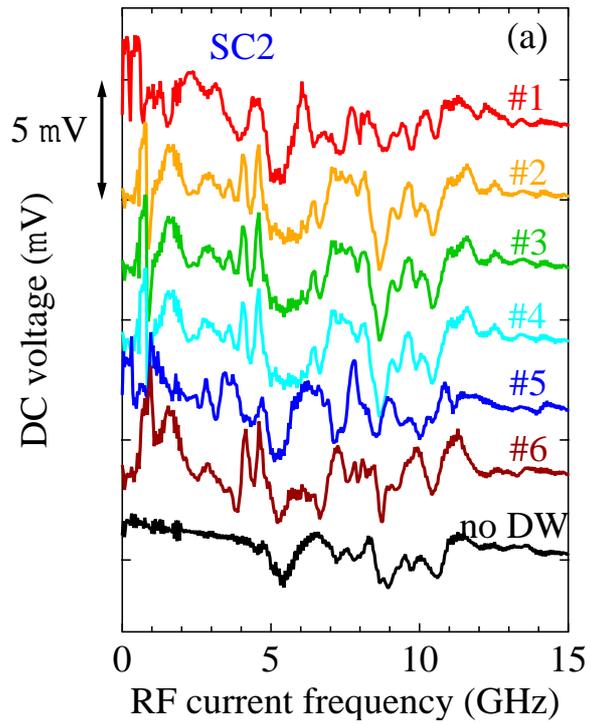

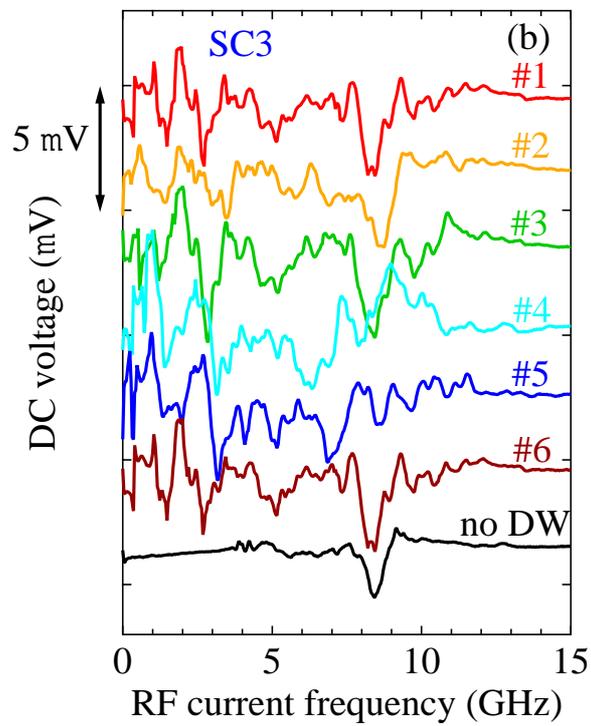

Fig. 4

Table I. Summary of wire shapes and sizes used in the present study

| Sample | Type | Width (nm) | Thickness (nm) | Radius (μm) |
|---|---|---|---|---|
| SC1 | Boomerang | 300 | 50 | 3 |
| SC2 | Semicircular | 300 | 30 | 10 |
| SC3 | Semicircular | 500 | 30 | 10 |
| SC4 | Semicircular | 1000 | 30 | 10 |